\documentclass[10pt,notitlepage]{iopart}
\usepackage{amssymb}
\usepackage{color}

\begin{document}
\title{Nondynamical modeling of resonances in a Quantum Field formalism {\rm\small(\today)}}

\author{Peter Morgan}
\address{Physics Department, Yale University, New Haven, CT 06520, USA.}
\ead{peter.w.morgan@yale.edu}
\date{\today}

\begin{abstract}
Well--defined nonlinear deformations of free quantum fields are introduced as manifestly Poincar\'e invariant scaling and resonance properties of non--dynamical scale models in Minkowski space, instead of introducing nonlinear dynamical deformations of free quantum fields that require the various truncations and scaling corrections of regularization and renormalization.
With the given algebraic construction, energy and momentum operators can be constructed \emph{ex post facto} as the generators of translations.
A weakened version of microcausality emerges naturally, ``convex hull microcausality'' ---that operators associated with two regions of space--time must commute if \emph{the convex hulls of} those regions are space--like separated---, which is enough for us to be able to construct an abundant class of interacting quantum fields.
\end{abstract}
\maketitle

\newcommand\Half{{\frac{1}{\rule{0pt}{1.4ex}2}}}
\newcommand\Intd{{\mathrm{d}}}
\newcommand\Remove[1]{{\raise 1.2ex\hbox{$\times$}\kern-0.8em \lower 0.35ex\hbox{$#1$}}}
\newcommand\SmallFrac[2]{{\scriptstyle\frac{\scriptstyle #1}{\scriptstyle#2}}}
\newcommand\Vacuum{{\left|\mathsf{V}\right>}}
\newcommand\dualVacuum{{\langle \mathsf{V}|}}
\newcommand\VEV[1]{{\left<\mathsf{V}\right|\hspace{-0.1em}#1\left|\mathsf{V}\right>}}
\newcommand\VacuumProjector{{\Vacuum\hspace{-0.1em}\dualVacuum}}
\newcommand\GibbsVacuum{{\left|\mathsf{G}\right>}}
\newcommand\GibbsVEV[1]{{\left<\mathsf{G}\right|\hspace{-0.1em}#1\left|\mathsf{G}\right>}}
\newcommand\eqdef{{\stackrel{\scriptstyle\mathrm{def}}{=}}}
\renewcommand\rme{{\mathrm{e}}}
\newcommand\rmj{{\mathsf{j}}}
\newcommand\NN{{{}^{{}_\mathcal{N}}\!}}
\newcommand\frameEQ[1]
        {{    \vspace{0.5ex}\newline\centerline{\fbox{$\displaystyle#1$}}\vspace{0.5ex}    }}
\newcommand\Unit{{\hat{\mathbf{1}} }}
\newcommand\oFG{{\frac{(g,\!f)(f,\!g)}{(g,\!g)(f,\!f)}}}

\newcommand\iD{{\mathrm{i}\hspace{-1.5pt}\Delta}}

\section{Introduction}
Scaling properties of a quantum field theory have become of paramount interest, in the form of the renormalization group\cite{WilsonKogut,Brown,Hollowood}, but in current methods they are applied essentially as an afterthought, after an ill--defined dynamics has first been decided upon.
Here, we will not approach dynamics until after we have considered how different scales are correlated with each other across space and time, which will be enough for us to construct a unitary evolution after that has been done. In not stating a Hamiltonian, but instead deriving the evolution after the fact, we will take these words of Wilson and Kogut\cite[p. 79]{WilsonKogut} with maximum seriousness: ``the behavior of the system is determined primarily by the fact that there is cooperative behavior, plus the nature of the degrees of freedom themselves. The interaction Hamiltonian plays only a secondary role.''
We will implement ``cooperative behavior" as different types of explicitly constructed well--defined stochastic resonances between different wave--numbers: not having a Hamiltonian will mean that we will not have to regularize or renormalize it.

Instead of working with a free quantum field operator--valued distribution $\hat\phi(x)$, we will work in the first instance with a quantum field operator formalism, which for the free field case can be constructed as $\hat\phi_f=\int\hat\phi(x)f(x)\Intd^4x$, using a \emph{modulation function} $f(x)$, so--called here because it describes how states and measurement results are \emph{modulated} when $\hat\phi_f$ and other operators are successively used to modulate a vacuum vector, $\Vacuum\mapsto\hat\phi_{f_1}\Vacuum$, $\Vacuum\mapsto\hat\phi_{f_1}\hat\phi_{f_2}\Vacuum$, and so on, where we take the vacuum vector to be a noisy version of a \emph{carrier signal} for the modulations $f_1$, $f_2$,\,...\,.
We will emphasize how using different modulation functions changes measurement results in response, where a precise model of a measurement is specified by giving some combination of modulation functions, together with how they are to be used as sums and products of operators $\hat\phi_{f_1}$, $\hat\phi_{f_2}$, ... .

Although the traditional construction $\hat\phi_f$ for the free quantum field is linear, we will here \emph{not} introduce nonlinearity by working with powers of the operator--valued distribution $\hat\phi(x)$, we will \emph{instead} introduce nonlinearity by working with operators $\hat\xi_f$ in a way that does not assume that $\hat\xi_f$ is a linear functional of $f$, so that in general $\hat\xi_{\lambda f+\mu g}\not=\lambda\hat\xi_f+\mu\hat\xi_g$.
To keep the discussion relatively simple, we will work here mostly with scalar quantum fields, so that modulation functions will be scalar functions, taken from a suitable function space, which we will take to be the Schwartz space of smooth functions for which the fourier transform exists and is also smooth, following the example of Haag\cite[\S II.1.2]{Haag}.

We will take it that if we can construct a useful class of \emph{interacting} quantum field operators $\hat\zeta_f$, for any modulation $f$, using only scaling and resonance properties, which we will in \S\ref{DeformationsII}, then we can define a unitary translation operator $\hat U(a)$ by its action $\hat U(a)\hat\zeta_f\hat U(a)^\dagger=\hat\zeta_{f_a}$, where $f_a(x)=f(x+a)$, and its trivial action on the translation invariant vacuum vector $\hat U(a)\Vacuum=\Vacuum$, from which we can derive a Hamiltonan dynamics, if we so wish.
The manifestly Poincar\'e invariant algebra implicitly contains all dynamical information.

Section \ref{RenormalizationNonlinearity} discusses in what ways nonlinearity of the quantum field is natural, particularly by reference to renormalization.
Section \ref{FreeKG} introduces the free Klein--Gordon quantum field and an elementary discussion of the traditional way in which the dynamics is deformed, with a marginally subversive discussion of that deformation given in \ref{OperatorCloud}, which shows that we can think of that traditional approach as constructing an ``operator--cloud'' that is associated with the backward light--cone of the support of a modulation function.
Section \ref{DeformationsI} introduces a class of nonlinear deformations that satisfy traditional microcausality, then Section \ref{ConvexHullMicrocausality} introduces a significantly different class of nonlinear deformations that do not satisfy microcausality, but do satisfy ``convex hull microcausality''.
Section \ref{DeformationsII} uses those new deformations to construct an operator--cloud that is associated with the regions of space--time that are obtained by repeated convolution of a modulation function with itself, but that nonetheless satisfies convex hull microcausality.
Section \ref{Multilinearity} shows that we can think of nonlinear quantum fields as a way to generate multilinear quantum fields, using polarization, with the reverse reconstruction of nonlinear quantum fields from those multilinear quantum fields being possible in elementary cases.
The construction by polarization of multilinear quantum fields is also notable for supporting an alternative motivation for nonlinearity, through consideration of Haag's theorem\cite{Fraser,EarmanFraser}.

It should be emphasized that this construction is just a first possibility for nonlinear constructions in a dual formalism that emphasizes modulation functions instead of working in a relatively ill--defined way with operator--valued distributions.
Working with scaling and resonance properties is very much in line with the understanding of renormalization that has been developed over many years, however other symmetries and other geometrical properties are likely to be more important if we consider either quantum field theory on a curved background or quantum gravity, and there may be other physically useful constructions at larger or smaller scales.

\section{Nonlinearity and renormalization}\label{RenormalizationNonlinearity}
In a classical context, we would not, except as a first approximation, expect a multiplicative modulation of a Gibbsian thermal vector state $\GibbsVacuum$, acting as a carrier for modulations, to obtain $\hat\xi_f\GibbsVacuum$, say, to result in a precisely linear response when we compare it to a different modulation, $\hat\xi_g\GibbsVacuum$: $\GibbsVEV{\hat\xi_g^\dagger\hat\xi_f}$ might or might not be an exactly linear functional of both $f$ and $g$.
Crucially, there are two linearities in quantum field theory: (1) the linearity of the vacuum and other states as maps from the $*$--algebra of operators that are used as models for measurements to expectation values, which for the vacuum state is $\hat A\mapsto\VEV{\hat A}$, which is required for there to be a probability interpretation of the theory and will therefore be carefully preserved; and (2) the linearity of the map of modulation functions into the $*$--algebra of operators, $f\mapsto\hat\xi_f$, for which there is no immediate necessity except convenience and tractability, so that it will be relinquished here.
We can also note, tangentially but somewhat in support, that axiomatic quantum field theory often introduces a unitary operator $\hat W_f=\rme^{\rmi\hat\phi_f}$ as a starting point\cite{BuchholzEM}, which for the free field case satisfies the algebra relation $\hat W_f\hat W_g=\hat W_{f+g}\rme^{-[(f^*,g)-(g^*,f)]/2}$ and which is clearly not a linear functional of $f$.

Renormalization gives \emph{some} motivation for introducing nonlinearity.
In Hollowood's account of renormalization group flow\cite[\S 1.2]{Hollowood}, for every physical measurement $F$ in a given experiment, depending on a length scale $\ell$, we want a theoretical model to be invariant under changes of the cutoff scale $\mu\rightarrow\mu'$,
\begin{equation}
  F(g_i(\mu);\ell)_\mu=F(g_i(\mu');\ell)_{\mu'},\qquad\mu,\mu'>\ell^{-1},
\end{equation}
where the $g_i(\mu)$ are the cutoff dependent couplings.
In a less idealized theoretical model, $\ell$ would be replaced by much more geometrical detail, which is what each modulation function is intended to provide, as a list of amplitudes $\tilde f(k)$ at each wave--number, so, replacing $\ell$ by a possibly very long list of modulation functions, we obtain
\begin{equation}
  F(g_i(\mu);f_1, ..., f_n)_\mu=F(g_i(\mu');f_1, ..., f_n)_{\mu'},\quad\mu,\mu'>\mathsf{AnInverseLengthDerivedFrom}(f_1, ..., f_n),
\end{equation}
where $\mathsf{AnInverseLengthDerivedFrom}(f_1, ..., f_n)$ in general will be mathematically rather arbitrary if we ever allow the modulation functions to be localized, because in that case the support of the fourier transforms of modulation functions would be unbounded.
In this construction $\mu$ and $\mu'$ are only weakly constrained by $f_1$, ..., $f_n$, however a more systematic approach, preferable just for being more systematic, would be to give the choice of $\mu$, $\mu'$ as different functionals of the modulation functions, $\mu(f_1, ..., f_n)$ and $\mu'(f_1, ..., f_n)$, giving
\begin{equation}
  F(g_i(\mu(f_1, ..., f_n));f_1, ..., f_n)_{\mu(f_1, ..., f_n)}
     =F(g_i(\mu'(f_1, ..., f_n));f_1, ..., f_n)_{\mu'(f_1, ..., f_n)},
\end{equation}
in which case the dependence of the couplings and of $F$ on $f_1$, ..., $f_n$ through $\mu$ and $\mu'$ (which are in any case too simplistic as single variables insofar as they are placeholders for \emph{all} the many choices made for a given systematic regularization scheme and renormalization scheme) can be rewritten as
\begin{equation}
  F(g_i(f_1, ..., f_n);f_1, ..., f_n)=F'(g_i'(f_1, ..., f_n);f_1, ..., f_n).\label{RGByModulation}
\end{equation}
There is no expectation that $F(g_i(\mu);\ell)_\mu$ is a linear function of $\ell$ or of the $g_i$, so there is no expectation that
\begin{equation}
  F(g_i(f_1, ..., f_n);f_1, ..., f_n)=F(\mathring g_i;f_1, ..., f_n),
\end{equation}
with the $\mathring g_i$ being constant coupling constants, must be a linear functional of $f_1$, ..., $f_n$, which at least leaves open the question of whether a quantum field operator $\hat\xi_f$ must be a linear functional of $f$.
If it is useful to take $\hat\xi_f$ to be a nonlinear functional of $f$, we might prefer to do so.

\section{The free Klein--Gordon quantum field and its traditional deformations}\label{FreeKG}
The aim of interacting quantum field theory has been to deform an operator--valued distribution such as the free Klein--Gordon quantum field, which satisfies
\begin{equation}\partial_\mu\partial^\mu\hat\phi(x)+m^2\hat\phi(x)=0,
\end{equation}
so that for the derivative $V'(\cdot)$ of a potential function $V(\cdot)$ an interacting quantum field $\hat\xi(x)$ satisfies
\begin{equation}\partial_\mu\partial^\mu\hat\xi(x)+m^2\hat\xi(x)+{:}V'(\hat\xi(x)){:}=0,
                           \label{UsualDeformation}
\end{equation}
where the normal--ordering $:\cdots:$ is an inadequate first step to make this latter equation well-defined.
If we work very formally and unrigorously, we can give, as a solution of this equation~\cite[\S 6-1-1]{IZ},
\begin{equation}\hat\xi(x)=\mathsf{T}^\dagger\!\left[\rme^{-\rmi\hat\mathcal{L}(x)}\right]\!
                     \hat\phi(x)  \mathsf{T}\!\left[\rme^{-\rmi\hat\mathcal{L}(x)}\right],\label{TraditionalSolution}
\end{equation}
where $\hat\mathcal{L}(x)=\int_{y_0<x_0}\!{:}V(\hat\phi(y)){:}\,\Intd^4y$ and the time--ordering $\mathsf{T}[\cdots]$ emphasize a dynamical approach to quantum field theory in contrast to a more structural approach that emphasizes scaling and resonance properties in Minkowski space.
\ref{OperatorCloud} shows, with a historically inevitable lack of rigor, that this can be thought of as equivalent to constructing an operator--cloud that is associated with the backward light--cone.

The free Klein--Gordon quantum field can alternatively be presented in a dual formalism as satisfying
\begin{equation}\hat\phi_{\partial_\mu\partial^\mu\!f+m^2\!f}=0,\mbox{ for any modulation function}\, f,\label{freeKG}
\end{equation}
because $\partial_\mu\partial^\mu$ is a self--adjoint operator.
The map $f\mapsto\partial_\mu\partial^\mu\!f+m^2\!f$ maps the modulation function $f$ to a function that has no component on the forward and backward mass--shells, however the vacuum state precisely acts to project to the forward mass--shell components.
For momentum space raising and lowering operator--valued distributions $a^\dagger(k)$ and $a(k)$ for the free Klein--Gordon quantum field, we have the standard Poincar\'e invariant commutator, in operator--valued distribution form\cite[Eq. 3-56]{IZ},
\begin{equation}[a(k),a^\dagger(k')]=\hbar(2\pi)^4\delta(k-k')2\pi\delta(k{\cdot}k-m^2)\theta(k_0),\label{kkCommutator}
\end{equation}
which becomes, for the lowering operator $a_f=\int a(k)\tilde f^*(k)\frac{\Intd^4k}{(2\pi)^4}$ and a conjugate raising operator $a_g^\dagger$, in operator form,
\begin{equation}[a_f,a^\dagger_g]=(f,g)=\hbar\!\int\!\tilde f^*(k)\tilde g(k)
                 2\pi\delta(k{\cdot}k-m^2)\theta(k_0)\frac{\Intd^4k}{(2\pi)^4},\label{IP}
\end{equation}
where $(f,g)$ is a pre--inner product on the modulation function space that is non--trivial only for modulation functions that have a component on the forward mass--shell.
Consequently, we have $(g,\partial_\mu\partial^\mu\!f+m^2\!f)=0$, trivial for all modulation functions $f$ and $g$, because $\partial_\mu\partial^\mu\!f+m^2\!f$ becomes $-(k{\cdot}k-m^2)\tilde f(k)$ as a fourier transform.
The raising and lowering operators allow us to construct a quantum field operator $\hat\phi_f=a_{f^*}+a_f^\dagger$, which solves Eq. (\ref{freeKG}) for any modulation function $f$, because in the vacuum state of the free Klein--Gordon quantum field $\VEV{\cdots\hat\phi_f\cdots}$ results in terms that all include a factor $(f^*,\cdot)$ or $(\cdot, f)$.
The pre--inner product $(f,g)$ satisfies microcausality because $[\hat\phi_f,\hat\phi_g]=(f^*,g)-(g^*,f)$ is zero if the modulation functions have supports that are space--like separated.
For a modulation function $f(x)$, $\tilde f(k)=\int f(x)\rme^{\rmi k{\cdot}x}\Intd^4x$, and similarly for $g(y)$,
\begin{equation}(f^*,g)-(g^*,f)=\hbar\int f(x)g(y)
                    \left[\rme^{\rmi k{\cdot}(y-x)}-\rme^{-\rmi k{\cdot}(y-x)}\right]
                    2\pi\delta(k{\cdot}k-m^2)\theta(k_0)\frac{\Intd^4k}{(2\pi)^4}\Intd x\Intd y,
\end{equation}
for which the integral with respect to $k$ is zero, by symmetry, provided the vector $y-x$ is space--like, which is satisfied if the modulation functions $f$ and $g$ have supports that are space--like separated.

\section{Deformations I}\label{DeformationsI}
Once we engage with a modulation function formalism, natural ideas of what a deformation might look like are rather different from Eq. (\ref{UsualDeformation}), such as
\begin{equation}\hat\xi_{\partial_\mu\partial^\mu\!f+m^2\!f+V'(f)}=0,
                                     \mbox{ for any modulation function}\, f,\label{V1}
\end{equation}
or
\begin{equation}\hat\xi_{\partial_\mu\partial^\mu\!f+m^2\!f}+V'(\hat\xi_f)=0,
                                     \mbox{ for any modulation function}\, f.\label{V2}
\end{equation}
Unlike deformations that are constructed using polynomials of operator--valued distributions, these and similar but more complicated equations are well-defined, although there is no guarantee that a particular such choice will have solutions, nor that any solutions will be useful physical models.

A well--known construction, called a generalized free field\cite[\S 3.4]{Streater}, uses raising and lowering operators that are non--trivial if $f$ has a component on \emph{either} of several forward mass--shells, with commutator
\begin{equation}[a_f,a^\dagger_g]=(f,g)_1+(f,g)_2+ \dots .
\end{equation}
Crucially, for any collection of modulation functions $f_1,..., f_n$, the Gram matrices $(f_i,f_j)_1$, $(f_i,f_j)_2$, ..., for each different mass $m_k$, and their sum $[a_{f_i},a_{f_j}^\dagger]=\sum_k(f_i,f_j)_k$ are all positive semi--definite matrices, which is enough to allow the construction of a Hilbert space for an arbitrary number of modulation functions, and hence for the construction of Fock space as an inductive limit, taking the induction to be over a Schwartz space of modulation functions on Minkowski space.
Note that for physics models in a manifestly Poincar\'e invariant formalism it is enough to introduce a finite number of modulation functions to model whatever finite experimental raw data we may have or anticipate being available, but for mathematical analysis it will be necessary for some purposes to take the inductive limit over the whole Schwartz space.
With the raising and lowering operators constructed in this way, the quantum field operator   $\hat\xi_f=a_{f^*}+a_f^\dagger$ is still a linear functional, $\hat\xi_{\lambda f+\mu g}=\lambda\hat\xi_f+\mu\hat\xi_g$, which satisfies the higher--order equation
\begin{equation}\hat\xi_{(\partial_\mu\partial^\mu+m_1^2)(\partial_\mu\partial^\mu+m_2^2)\cdots f}=0,\mbox{ for any modulation function}\, f.
\end{equation}

With the dropping of linearity as a requirement, we can also introduce raising and lowering operators that are non--trivial if $f$ has a component on \emph{all} of several forward mass--shells, with commutator
\begin{equation}[a_f,a^\dagger_g]=(f,g)_1(f,g)_2 \cdots .
\end{equation}
The Hadamard product $[a_{f_i},a_{f_j}^\dagger]=(f_i,f_j)_1(f_i,f_j)_2\cdots$ is, as required, a positive semi--definite matrix\footnote{For a Hadamard product $M_{ij}=A_{ij}B_{ij}$, where $A$ and $B$ are both positive semi--definite matrices, so that they can be written as sums $A_{ij}=\sum_k U^*_{ki}U_{kj}$ and $B_{ij}=\sum_k V^*_{ki}V_{kj}$, $M_{ij}=\sum_k\sum_\ell U^*_{ki}V^*_{\ell i}V_{\ell j}U_{kj}=\sum_{k\ell}W^*_{k\ell i}W_{k\ell j}$, where $W_{k\ell i}=U^*_{ki}V^*_{\ell i}$, so that $M$ is also a positive semi--definite matrix.}, which is again enough to allow the construction of a Hilbert space.
$\hat\xi_f=a_{f^*}+a_f^\dagger$ satisfies multiple differential equations,
\begin{equation}\hat\xi_{\partial_\mu\partial^\mu\!f+m_1^2f}=0\mbox{\quad\sl and\quad}
                       \hat\xi_{\partial_\mu\partial^\mu\!f+m_2^2f}=0\mbox{\quad\sl and\quad}\cdots,
                               \mbox{ for any modulation function}\, f,
\end{equation}
and microcausality is satisfied because $(f^*,g)_k=(g^*,f)_k$ for every $k$, but $\hat\xi_f$ is not a linear functional, $\hat\xi_{\lambda f+\mu g}\not=\lambda\hat\xi_f+\mu\hat\xi_g$.
The spectrum of 4--momentum measurements is confined to the forward light--cone because $(f,g)_k$ projects to the forward light--cone for every $k$.
The two approaches can be combined to give an arbitrary sum of products of this kind, so that in general the commutator can be a sum of products for many different masses,
$[a_f,a^\dagger_g]=\sum_k\prod_\ell(f,g)_{k\ell}$.

\newcommand\xiN[2]{{\hat\xi^{\hspace{-0.07em}{\star}\hspace{-0.07em}#1}_{#2}}}
\newcommand\aN[2]{{a^{\hspace{-0.1em}{\star}\hspace{-0.07em}#1}_{#2}}}
\section{Convex Hull Microcausality}\label{ConvexHullMicrocausality}
Convex hull microcausality is a less constrained microcausality, which requires only that the modulation functions have \emph{the convex hulls of} their supports space--like separated.
Historically, this is somewhat presaged by the effective field physics of quasi--particles and collective excitations, which work with the center of mass or otherwise positively weighted centers of many interacting components, for which such centers may fall outside the supports of any of the components but will be within the convex hull of the union of those supports.

As a simplest example, convex hull microcausality is satisfied by a quantum field for which
\begin{eqnarray}
  [\xiN{2}{f},\xiN{2}{g}]=([f{\star}f]^*,[g{\star}g])-([g{\star}g]^*,[f{\star}f]),
\end{eqnarray}
using convolution, for which $\widetilde{[f{\star}f]}(k)=\tilde f^2(k)$,
\begin{eqnarray}([f{\star}f]^*,[g{\star}g])&{-}&([g{\star}g]^*,[f{\star}f])=\hbar\!\int\!
                    \left[\rme^{\rmi k{\cdot}((y_1+y_2)-(x_1+x_2))}
                            -\rme^{-\rmi k{\cdot}((y_1+y_2)-(x_1+x_2))}\right]\cr
                    &&\hspace{1em}\times\ 2\pi\delta(k{\cdot}k-m^2)\theta(k_0)\frac{\Intd^4k}{(2\pi)^4}
                     f(x_1)f(x_2)g(y_1)g(y_2)\Intd x_1\Intd x_2\Intd y_1\Intd y_2,
\end{eqnarray}
because $(x_1+x_2)/2$ and $(y_1+y_2)/2$ are always within the convex hull of the supports of $f$ and $g$ respectively, so $(y_1+y_2)-(x_1+x_2)$ is space--like and convex hull microcausality is satisfied if the convex hulls of the supports of $f$ and $g$ are space-like separated.
Convex hull microcausality is easily seen also to be satisfied if we use $f^{\star n}$, for which $\widetilde{[f^{\star n}]}(k)=\tilde f^n(k)$, instead of the simplest form, $f^{\star 2}=f{\star}f$, in the above.

This is a little surprising, because the support of $f^{\star n}$ will overlap the support of $g^{\star n}$ for large enough $n$, but symmetry ensures that $(f^{\star n*},g^{\star n})-(g^{\star n*},f^{\star n})$ nonetheless is zero if the convex hulls of the supports of $f$ and $g$ are space-like separated (just as symmetry ensures that microcausality is satisfied for interacting quantum fields despite the backward light cone operator--clouds overlapping).
It is possible that other constructions might have a different space--time symmetry that satisfy convex hull or other modifications of microcausality, but we will here pursue only constructions of this general form.
With this property defined, we can introduce raising and lowering operators such as
\begin{equation}[\aN{n}{f},\aN{n}{g}^\dagger]=\delta_{m,n}(f^{\star n},g^{\star n}),
\end{equation}
for which the fields $\xiN{n}{f}=\aN{n}{f^*}+\aN{n}{f}^\dagger$ independently satisfy convex hull microcausality for different $n$.

We can generalize this construction by introducing a scaled modulation function, for which, for $\alpha>0$, \begin{equation}\widetilde{f^{(\alpha)}}(k)=\int f(x)\rme^{\rmi\alpha k{\cdot}x}\Intd^4x
       =\int f(\SmallFrac{x}{\rule{0ex}{1.15ex}\alpha})\rme^{\rmi k{\cdot}x}\Intd^4x,
\end{equation}
and a convolution of $\mathsf{n}$ modulation functions, scaled by $\alpha_1$, ..., $\alpha_\mathsf{n}$, a vector denoted by $\underline{\alpha}$,
\begin{equation}\widetilde{f^{(\underline{\alpha})}}(k)
         =\prod_{j=1}^{\mathsf{n}}\widetilde{f^{(\alpha_j)}}(k)
         =\int \prod_{j=1}^{\mathsf{n}}\Bigl[f(x_j)\Intd^4x_j\Bigr]
                 \rme^{\rmi k{\cdot}(\sum_{j=1}^{\mathsf{n}}\alpha_j x_j)}.
\end{equation}
Writing $|\underline{\alpha}|=\sum_{j=1}^{\mathsf{n}}\alpha_j$, we can construct raising and lowering operators for which
\begin{equation}[a_f^{(\underline{\alpha})},a_g^{(\underline{\beta})\dagger}]
       =\delta_{|\underline{\alpha}|,|\underline{\beta}|}\,(f^{(\underline{\alpha})},g^{(\underline{\beta})}),
\end{equation}
where $\delta_{|\underline{\alpha}|,|\underline{\beta}|}$ is 1 if $|\underline{\alpha}|=|\underline{\beta}|$ or otherwise is 0, for which the quantum fields
$\hat\xi_f^{(\underline{\alpha})}=a_{f^*}^{(\underline{\alpha})}+a_f^{(\underline{\alpha})\dagger}$
satisfy convex hull microcausality,
\begin{eqnarray}[\hat\xi_f^{(\underline{\alpha})},\hat\xi_g^{(\underline{\beta})\dagger}]
      &=&[a_{f^*}^{(\underline{\alpha})},a_g^{(\underline{\beta})\dagger}]
                           -[a_{g^*}^{(\underline{\beta})},a_f^{(\underline{\alpha})\dagger}]\cr
      &=&\hbar\delta_{|\underline{\alpha}|,|\underline{\beta}|}\!\int\!
                     \prod_j\Bigl[f(x_j)\Intd^4x_j\Bigr]\prod_k\Bigl[g(y_k)\Intd^4y_k\Bigr]
                     2\pi\delta(k{\cdot}k-m^2)\theta(k_0)\frac{\Intd^4k}{(2\pi)^4}\cr
      &&\times\ \Biggl[\exp{\Biggl(\rmi k{\cdot}\Bigl(\sum_j\alpha_jx_j-\sum_k\beta_ky_k\Bigr)\Biggr)}
                  -\exp{\Biggl(-\rmi k{\cdot}\Bigl(\sum_j\alpha_jx_j-\sum_k\beta_ky_k\Bigr)\Biggr)}\Biggr],
\end{eqnarray}
because $|\underline{\alpha}|$ is required to be the same as $|\underline{\beta}|$, and $\sum_j\alpha_jx_j/|\underline{\alpha}|$ and $\sum_k\beta_ky_k/|\underline{\beta}|$ are within the convex hulls of the test functions $f$ and $g$ respectively, so their difference is a space--like 4--vector, and the integral over $k$ vanishes by symmetry if the convex hulls of the supports of $f$ and $g$ are space--like separated.

Finally, for a scaling vector $\alpha_1$, ..., $\alpha_\mathsf{n}$ we can introduce a local nonlinear functional $F[f](x_1,...,x_\mathsf{n})$, where $\mathsf{Supp}(F[f])\subseteq\mathsf{Supp}(f^{\times\mathsf{n}})$, allowing the construction of a quantum field $\hat\xi_f^{(F,\underline{\alpha})}=a_{f^*}^{(F,\underline{\alpha})}+a_f^{(F,\underline{\alpha})\dagger}$,
\begin{eqnarray}\hspace{-1em}[\hat\xi_f^{(F,\underline{\alpha})},\hat\xi_g^{(G,\underline{\beta})\dagger}]
      &=&[a_{f^*}^{(F,\underline{\alpha})},a_g^{(G,\underline{\beta})\dagger}]
                           -[a_{g^*}^{(G,\underline{\beta})},a_f^{(F,\underline{\alpha})\dagger}]\cr
      &=&\hbar\delta_{|\underline{\alpha}|,|\underline{\beta}|}\!\int\!
                     F[f](x_1,...\,)G[g](y_1,...\,)\prod_j\Bigl[\Intd^4x_j\Bigr]\prod_k\Bigl[\Intd^4y_k\Bigr]
                     2\pi\delta(k{\cdot}k-m^2)\theta(k_0)\frac{\Intd^4k}{(2\pi)^4}\cr
      &&\times\ \Biggl[\exp{\Biggl(\rmi k{\cdot}\Bigl(\sum_j\alpha_jx_j-\sum_k\beta_ky_k\Bigr)\Biggr)}
                  -\exp{\Biggl(-\rmi k{\cdot}\Bigl(\sum_j\alpha_jx_j-\sum_k\beta_ky_k\Bigr)\Biggr)}\Biggr],
\end{eqnarray}
which still satisfies convex hull microcausality. As well as polynomials, the local nonlinear functional $F[f]$ may include scalar nonlinear derivations such as $\frac{\partial f(x)}{\partial x^\mu}\frac{\partial f(x)}{\partial x_\mu}$.

These fields will give us a well--defined way to construct convex hull microcausal deformations as an operator--cloud that is associated with all space--time, instead of introducing time--ordering and ill--defined Lagrangian interaction terms.

\section{Deformations II}\label{DeformationsII}
In Sections \ref{DeformationsI} and \ref{ConvexHullMicrocausality}, the constructions use only raising and lowering operators, so although the induced evolutions of measurement results will exhibit considerably different geometries, the vacuum state and coherent states ---eigenstates of the lowering operators--- are straightforwardly Gaussian, just as for the free field.
For the traditional deformation, Eq. (\ref{UsualDeformation}), in contrast, the vacuum state is not Gaussian because it constructs an operator--cloud in the backward light--cone as its lowest--level measurement operation.
Given Sections \ref{DeformationsI} and \ref{ConvexHullMicrocausality}, an obvious way to leverage convex hull microcausality is to construct an operator--cloud in the regions of space--time that result from convolutions such as $\hat\xi_f^{(\underline{\alpha})}$ for various $\underline{\alpha}$, where $\hat\xi_f=\hat\xi_f^{(1)}$.
For a Hermitian operator $\hat\mathcal{S}_f$ that uses such convolutions, we can, for example, construct a quantum field
\begin{equation}
  \hat\zeta_f=\rme^{+\rmi\hat\mathcal{S}_f}\hat\xi_f\rme^{-\rmi\hat\mathcal{S}_f}.
\end{equation}
We will introduce several ways to construct $\hat\mathcal{S}_f$, of increasing complexity, by example:
\begin{itemize}
\item
{\large$\displaystyle\hat\mathcal{S}_f\,{=}\,\sum_{n=1}^N\lambda_n
                  \hat\xi_f^{(\underline{\alpha}_n)}{}^\dagger\hat\xi_f^{(\underline{\alpha}_n)}
                  \hat\xi_f^{(\underline{\alpha}'_n)}{}^\dagger\hat\xi_f^{(\underline{\alpha}'_n)}$}, where
$|\underline{\alpha}_n|\not=|\underline{\alpha}'_n|$ so that $\hat\mathcal{S}_f^\dagger=\hat\mathcal{S}_f$, which constructs an operator--cloud associated with the space--time region associated with the highest convolution power $\underline{\alpha}_n$ or $\underline{\alpha}'_n$.
Note that if for all $n$, $|\underline{\alpha}_n|\not=1$ and $|\underline{\alpha}'_n|\not=1$, then this construction will be trivial because in that case every term would commute with $\hat\xi_f$; similarly, each added term must not commute with at least one other previously added term.
Each successive level of approximation introduces resonances associated with different scales, either smaller or larger.
The sum over $n$ for the chosen values of $\underline{\alpha}_n$ and $\underline{\alpha}'_n$ and for an appropriate choice of modulation function $f$ is comparable to the integration over all points in the backward light--cone.\newline
With this construction, however, we only introduce resonances at wave--numbers that are multiples of wave--numbers for which $\tilde f(k)\not=0$, which the next construction goes beyond.

\item
{\large$\displaystyle\hat\mathcal{S}_f\,{=}\,\sum_{n=1}^N\lambda_n
        \hat\xi_f^{(F_n,\underline{\alpha}_n)}{}^\dagger\hat\xi_f^{(F_n,\underline{\alpha}_n)}
        \hat\xi_f^{(F'_n,\underline{\alpha}'_n)}{}^\dagger\hat\xi_f^{(F'_n,\underline{\alpha}'_n)}$}.
If we use the simplest local nonlinear functional, $F_n[f](x_1,...\,)=\prod_j [f(x_j)]^2$, for example, for which the fourier transform of each factor is $\widetilde{f^2}(k_j)=\int\tilde f(u_j)\tilde f(k_j-u_j)\SmallFrac{\Intd^4k_j}{\rule{0pt}{1.75ex}(2\pi)^4}$, then we can obtain nontrivial resonances even when there are no wave--numbers for which $\tilde f(k)$ and $\tilde g(k)$ are both nonzero when we consider transition amplitudes such as, as the most elementary case,
$|\VEV{\hat\zeta_g^\dagger\hat\zeta_f}|^2$, giving a nontrivial theory because the interactions cannot be diagonalized in momentum space (see, for example, \cite[p.85]{WilsonKogut}).

\item
We can use an arbitrary product of an arbitrary number of different {\large$\displaystyle\hat\xi_f^{(F_n,\underline{\alpha}_n)}$} for each term in the sum. There is no need for $\hat\mathcal{S}_f$ to be positive semi--definite because the confinement of the spectrum of 4--momentum measurements to the forward light--cone is independently ensured by the projection to the forward light--cone that is required of all of the pre--inner products $(f,g)_i$ that are used.

\item
We can replace the $\sum_{n=1}^N$ by an integration over a range of values $\nu$, provided there is only a countable number of different values for $|\underline{\alpha}_\nu|$, $|\underline{\alpha}'_\nu|$, \emph{et cetera}: without this constraint, we would again find ourselves multiplying distributions.
\end{itemize}
It should not be thought surprising that there are at least this many options, because the mathematics of quantum field theory fixes a system of probability densities over a field of observables, which is a significantly higher--order mathematical structure than a classical field, and the regularization schemes and renormalization schemes of traditional interacting quantum fields also conceal much complexity.
With this enormous flexibility, it will be hard to prove that no possible choice of such structures and of modulation functions can model a given collection of experiments, although we may find that such models are too intractably complex for them to be useful.

The symmetries of the expression for quantum field operators $\hat\zeta_f$ will result in the $*$--algebra $\mathcal{A}_\zeta$ that is generated by them being the same as or a sub--$*$--algebra of the $*$--algebra $\mathcal{A}_\xi$ that is generated by all the quantum field operators $\hat\xi_f^{(\underline{\alpha})}$ that are used in the construction of the $\hat\zeta_f$.
If $\mathcal{A}_\zeta=\mathcal{A}_\xi$, then it becomes a delicate question whether one or the other is more important, however for most interesting cases we would expect $\mathcal{A}_\zeta\subset\mathcal{A}_\xi$, as a nontrivial consequence of the symmetries of specific stochastic resonance conditions.

\section{Polarization and Multilinearity}\label{Multilinearity}
There is another way to motivate the above discussion, which was introduced in \cite{Morgan1507} and which some may find more compelling.
Given a nonlinear quantum field $\hat\zeta_f$, one possibility is to think of it as a generating function from which we can construct a system of symmetric multilinear quantum fields, each of which is linear in each of $n$ modulation functions $f_1$, ..., $f_n$, by polarization,
\begin{equation}
  \VEV{\hat B^\dagger\hat\zeta^{(n)}_{f_1,...,f_n}\hat A}
      =\frac{1}{n!}\left.\prod_{j=1}^n\left[\frac{\partial}{\partial\lambda_j}\right]
           \VEV{\hat B^\dagger\hat\zeta_{\sum_{k=1}^n\lambda_kf_k}\hat A}
                    \right|_{\mbox{\small$\lambda_1= ... =\lambda_n=0$}},
\end{equation}
from which we can construct a representation of a quantum field $\hat\zeta(x_1, ..., x_n)$ that is a symmetric multi--point operator--valued distribution.
Reconstruction of $\hat\zeta_f$ if we are given $\hat\zeta^{(n)}(x_1, ..., x_n)$ for all $n$ will be possible if $\hat\zeta^{(n)}(x_1, ..., x_n)\not=0$ only for a finite number of such $n$, but will in general not be possible. Whether $\hat\zeta_f$ can be reconstructed or not, we can use $\hat\zeta^{(n)}_{f_1,...,f_n}$ and $\hat\zeta^{(n)}(x_1, ..., x_n)$ to construct models.
\cite{Morgan1507} suggests that
a possible reaction to Haag's theorem\cite{Fraser,EarmanFraser} is to think that the Fock--Hilbert space generated by free 1--particle Wightman fields\cite[Ch. II]{Haag} is not ``big enough'' to model bound states or interacting states generally, which can be taken to suggest adding multi--particle bound and interacting states explicitly as multi--point fields $\hat\phi(x_1, ..., x_n)$ and constructing physical states as superpositions of different products of 1--particle and multi--particle fields acting on the vacuum state.
We are already accustomed to introducing quasiparticles and collective excitations that are effectively bound multi--particle fields, such as phonons in statistical physics and such as for protons in high energy physics, and to using a simple center--of--mass wave function as a first approximation for the interference patterns that can be generated using large molecules.
From the point of view of \emph{this} paper, in contrast to \cite{Morgan1507}, that approach is less well--motivated than working with the nonlinear quantum field $\hat\zeta_f$, then perhaps thinking of it as a generating function for operator--valued distributions $\hat\zeta^{(n)}(x_1, ..., x_n)$, however some aspects of the construction are different depending on the starting point one adopts: in particular, one can construct multi--point fields that are not symmetric in their arguments.
In either case, however, the already disconcerting nonlocality of free quantum fields is only made greater by the introduction of nonlinearity and of convolution, though within the constraint of convex hull microcausality.

\section{Discussion}
As mentioned in the introduction, it should be emphasized that this construction is just a first possibility for nonlinear constructions in a dual formalism that emphasizes the modulation functions instead of working in an ill--defined way with operator--valued distributions.

The constructions of Sections \ref{ConvexHullMicrocausality} and \ref{DeformationsII} are integral forms that have not been and do not seem to be easily specified as solutions of differential equations.
In practice, however, effective quantum fields already only take a differential equation as a starting point, with higher--order and higher--degree terms introduced as necessary without much regard to what differential equation the resulting construction might satisfy.

The focus on nonlinearity and on scaling and resonance properties are both natural from a classical perspective, however each introduces a new form of nonlocality.
Nonlinearity, by polarization, introduces a linear multipoint structure that can be arbitrarily nonlocal, but still may satisfy microcausality in the sense that $[\hat\zeta_f,\hat\zeta_g]=0$ whenever the supports of $f$ and $g$ are space--like separated.
Resonance through the convolution pathway of Section \ref{ConvexHullMicrocausality} is also nonlinear, so it also introduces a linear multipoint structure, however it only satisfies convex hull microcausality, that $[\hat\zeta_f,\hat\zeta_g]=0$ whenever the convex hulls of the supports of $f$ and $g$ are space--like separated.
These introductions, however, can and should be seen as not significantly more global than the restriction to positive frequency that is introduced by free quantum fields: the differential equation $\partial_\mu\partial^\mu\hat\phi(x)+m^2\hat\phi(x)=0$ is purely local, but the use of the solution $2\pi\delta(k{\cdot}k-m^2)\theta(k_0)$ in Eq. (\ref{kkCommutator}) requires an essentially nonlocal choice of boundary conditions.

Nonlinearity may in some cases require us to be very careful when accurately modeling complex experiments.
The exact form of a low--frequency envelope applied to an otherwise high--frequency modulation may matter for some kinds of resonances, for example, even when the low and high frequency components are separated by many orders; similarly, the immediate harmonics of a modulation may make a very significant difference, so that a triangle waveform modulation may behave very differently from a sinusoidal modulation.
This sensitivity to both low-- and high--frequency components is rather contrary to the principles of the renormalization group, which looks for invariance of observables under scale transformations at large and small scales, however it is desirable to measure and to model dependencies on all features of any given modulation.
The choice of a detailed input modulation to achieve a desired response is already a significant task in classical nonlinear signal processing, which is made more difficult by the effects of quantum and other noise and the need to engineer probability measures for samples of a noisy signal, not just amplitudes of a signal, all of which requires even greater ingenuity.

A resonance in classical mechanics is a preferred frequency, so that if a system is externally driven by a signal that has components at that frequency, the system will respond with steadily increasing amplitude, until nonlinearity and other effects limit that increase.
For quantum fields, an applied modulation will cause, at other times and places, a response (which we can take to be a probability density for a sample space of possible measurement results) that depends on whether the measurement operator is \emph{tuned} to that applied modulation.
The mathematics above says that the applied modulation doesn't only drive its own wavenumbers, it also drives overtones, undertones, and more general auxiliary wavenumbers in ways that are quite analogous to the elaborate relationships between different modes of higher--dimensional systems such as a drum.
We can loosely think of that part of the structure of space-time over which we have no control as of the same type as boundary and material conditions of a drum (taking it that, for the sake of this comparison, we have no control of the boundary and materials of a drum for the duration of an experiment), insofar as whatever the structure of space--time might be controls what resonances are more or less important.
Without knowing anything about the structure of space--time, or of the boundary and material of a drum, we can describe the response, for a given measurement, to a given applied modulation.
To carry the analogy a little more, the sound of a drum ---the nonlinear response of our ears and our neural processing to drumsticks brushing or beating on a drumhead, or the look of the Wigner distribution functions of the voltages from differently placed microphones--- is ``shaped'' not only by the geometry of its boundary but also by its density, thickness, stiffness, tension, every detail of the crystalline and amorphous structure of all its parts, and by how it is mounted.
In abstract terms, all such boundary and material details contribute to the conditioning of measurement results and how they can most fruitfully be analyzed, however the finite numbers of measurements we make will not absolutely fix every detail of how the drum was made, and similarly we can expect measurements of the response to however we brush or beat or modulate the vacuum state not to fix its every detail.
We can, however, as empiricists, choose to deny the analogy with a drum altogether, so that we insist that there is \emph{only} the state, which describes measurement results, and that we can only tendentiously claim that there really is a {\sl carrier} of the state, which would only in overactive imagination {\sl explain} those measurement results.

A quantum field theory has historically told us a dynamics, which determines resonances as a model for what we observe, but if we approach the question less ambitiously, it can be useful and satisfying enough for a quantum resonance theory to model resonances directly.
The formalism introduced here may not be enough to model all the resonances we observe, but it is hopefully enough to begin a worthwhile new approach to the question.

I am grateful for comments from Stephen Paul King.

\appendix
\section{Interacting quantum fields as an operator--cloud}\label{OperatorCloud}
The argument presented here appeared in arxiv.org:1211.2831, which, however, is otherwise deservedly unpublished (so it is not included in the bibliography).
Superficially, the argument does little more than establish that Eq. (\ref{TraditionalSolution}) is a plausible solution for Eq. (\ref{UsualDeformation}), if we content ourselves with mathematics that is not well--defined, however it also establishes the less common idea that Eq. (\ref{TraditionalSolution}) can be understood to describe a structured cloud of operators associated with the backward light--cone of the free field operator--valued distribution that seeds the construction.

The components of $\hat\mathcal{L}(x)=\int_{y_0<x_0} V(\hat\phi(y))\Intd^4y$ (which is used in Eq. (\ref{TraditionalSolution})) that are space-like separated from $x$ commute with $\hat\phi(x)$, and because of time-ordering those components cancel with the time--reversed components of the inverse, so that we can substitute
\begin{equation}
  \hat\mathcal{L}(x)=\int\limits_{\blacktriangle(x)}\!\hat V(\hat\phi(y))\Intd^4y,
\end{equation}
where  $\blacktriangle(x)=\{y:(x-y)^2\ge 0\ \mathrm{and}\ y_0<x_0\}$ is the region of space--time that is light--like or time--like separated from and earlier than $x$.
Furthermore, the adjoint action of $\hat\phi(x)$ on a time-ordered expression is a derivation, because time-ordering ensures commutativity, so that, taking a quartic scalar interaction with Hamiltonian density $\SmallFrac{\lambda}{4!}\!:\!\hat\phi^4(y)\!:$ as an example,
\begin{equation}
  \left[\hat\phi(x),\mathsf{T}\!\left[\left(\int\!:\!\hat\phi^4(y)\!:\Intd^4y\right)^n\right]\right]=
       \mathsf{T}\!\left[\int\!4n\iD(x-z):\!\hat\phi^3(z)\!:\Intd^4z\left(\int\!:\!\hat\phi^4(y)\!:\Intd^4y\right)^{n-1}\right],
\end{equation}
where
\begin{equation}\label{iDG}
  \iD(x-z)=-\rmi(G_{\mathrm{ret}}(x-z)-G_{\mathrm{adv}}(x-z))=[\hat\phi(x),\hat\phi(z)]
\end{equation}
is the free field commutator and $G_{\mathrm{ret}}(x-z)$ and $G_{\mathrm{adv}}(x-z)$ are the retarded and advanced Green functions~\cite[\S 1-3-1]{IZ}.
For the interacting field, therefore, we have the construction
\begin{eqnarray}
  \hat\xi(x)&=&\mathsf{T}^\dagger\!\left[\rme^{-\rmi\hat\mathcal{L}(x)}\right]\hat\phi(x)
                          \mathsf{T}\!\left[\rme^{-\rmi\hat\mathcal{L}(x)}\right]\cr
            &&\hspace{3em}\mbox{[$\hat\phi(x)$ acts as a derivation, ...]}\cr
            &=&\mathsf{T}^\dagger\!\left[\rme^{-\rmi\hat\mathcal{L}(x)}\right]
                       \left(\!\mathsf{T}\left[\rme^{-\rmi\hat\mathcal{L}(x)}\right]\hat\phi(x)
                          -\mathsf{T}\!\left[\SmallFrac{\mathrm{i\lambda}}{3!}\!\!\!
                              \int\limits_{\blacktriangle(x)}\!\iD(x-z):\!\hat\phi^3(z)\!:\Intd^4z
                                    \rme^{-\rmi\hat\mathcal{L}(x)}\right]\right)\cr
            &=&\hat\phi(x)-\mathsf{T}^\dagger\!\left[\rme^{-\rmi\hat\mathcal{L}(x)}\right]
                          \mathsf{T}\!\left[\SmallFrac{\mathrm{i\lambda}}{3!}\!\!\!
                              \int\limits_{\blacktriangle(x)}\!\iD(x-z):\!\hat\phi^3(z)\!:\Intd^4z
                                    \rme^{-\rmi\hat\mathcal{L}(x)}\right]\cr
            &=&\hat\phi(x)-\SmallFrac{\mathrm{i\lambda}}{3!}\!\!\!\int\limits_{\blacktriangle(x)}\iD(x-z)
                                \mathsf{T}^\dagger\!\left[\rme^{-\rmi\hat\mathcal{L}(x)}\right]
                                \mathsf{T}\!\left[:\!\hat\phi^3(z)\!:\rme^{-\rmi\hat\mathcal{L}(x)}\right]\Intd^4z\cr
            &&\hspace{3em}
           \mbox{[components of $\hat\mathcal{L}(x)$ that are space--like separated from $z$}\cr
            &&\hspace{13em}
           \mbox{or are later than $z$ \emph{cancel}, leaving $\hat\mathcal{L}(z)$, ...]}\cr
            &=&\hat\phi(x)-\SmallFrac{\mathrm{i\lambda}}{3!}\!\!\!\int\limits_{\blacktriangle(x)}\iD(x-z)
                                \mathsf{T}^\dagger\!\left[\rme^{-\rmi\hat\mathcal{L}(z)}\right]
                                     :\!\hat\phi^3(z)\!:
                                \mathsf{T}\!\left[\rme^{-\rmi\hat\mathcal{L}(z)}\right]\Intd^4z\cr
            &=&\hat\phi(x)-\SmallFrac{\mathrm{i\lambda}}{3!}\!\!\!
                              \int\limits_{\blacktriangle(x)}\!\iD(x-z):\!\hat\xi^3(z)\!:\Intd^4z\quad
        \biggl[:\!\hat\xi^3(z)\!: \;=\mathsf{T}^\dagger\!\left[\rme^{-\rmi\hat\mathcal{L}(z)}\right]
                                     :\!\hat\phi^3(z)\!:
                                \mathsf{T}\!\left[\rme^{-\rmi\hat\mathcal{L}(z)}\right]\biggr]\cr
            &=&\hat\phi(x)-\SmallFrac{\lambda}{3!}\!\!
                              \int\!G_{\mathrm{ret}}(x-z):\!\hat\xi^3(z)\!:\Intd^4z,\label{IntegralEqn}
\end{eqnarray}
where the restriction to the backward light--cone $\blacktriangle(x)$ is equivalent to replacing the propagator $\iD(x-z)$ by $-\rmi G_{\mathrm{ret}}(x-z)$, as we see from Eq. (\ref{iDG}).
Insofar as we can take $:\!\hat\phi^3(z)\!:$ formally to be an infinite multiple of $\hat\phi(z)$ subtracted
from $\hat\phi^3(z)$, we can take $:\!\hat\xi^3(z)\!:$ formally to be the same infinite multiple of $\hat\xi(z)$ subtracted from $\hat\xi^3(z)$.
$\hat\phi(x)$ satisfies the homogeneous Klein--Gordon equation, $(\partial_\mu\partial^\mu+m^2)\hat\phi(x)=0$, and
$G_{\mathrm{ret}}(x-z)$ satisfies $(\partial_\mu\partial^\mu+m^2)G_{\mathrm{ret}}(x-z)=\delta^4(x-z)$, so, applying the operator
$(\partial_\mu\partial^\mu+m^2)$ to the last line of Eq. (\ref{IntegralEqn}), $\hat\xi(x)$ satisfies the
nonlinear differential equation
\begin{eqnarray}\label{FieldDiffEqn}
  (\partial_\mu\partial^\mu+m^2)\hat\xi(x)+\SmallFrac{\lambda}{3!}:\!\hat\xi^3(x)\!:\;=0.
\end{eqnarray}
The above construction shows that, apart from the mathematically ill--defined necessity to regularize and renormalize, we can construct an interacting field by replacing $\hat\phi(x)$ at a point by an operator--cloud at points of the backward light--cone of $x$, constructed using the propagator $G_{\mathrm{ret}}(x-z)$.
This is compatible with the usual, more practical use of the Feynman propagator and integration over all space--time, but it is worthwhile to recognize that the same algebraic structure can be presented using the retarded propagator, without introducing time--ordering, and with integration restricted to the backward light--cone.
We can also add an arbitrary solution $I(x-z)$ of the Klein--Gordon equation to the last line of Eq. (\ref{IntegralEqn}), to obtain
\begin{equation}
\hat\xi(x)=\hat\phi(x)-\SmallFrac{\lambda}{3!}\!\!
                              \int\!\Bigl[G_{\mathrm{ret}}(x-z)+I(x-z)\Bigr]:\!\hat\xi^3(z)\!:\Intd^4z,
\end{equation}
which allows us, as two special cases, to replace $G_{\mathrm{ret}}(x-z)$ by $G_{\mathrm{adv}}(x-z)$ or by $\Half\bigl[G_{\mathrm{ret}}(x-z){+}G_{\mathrm{adv}}(x-z)\bigr]$, thereby constructing operator--clouds that have the same local structure, satisfying Eq. (\ref{FieldDiffEqn}), but that are associated with different space--time regions, which, tendentiously, we can take as encouragement for the construction in the main text of operator--clouds associated with other space--time regions.


\vspace{4ex}


\begin{thebibliography}{9}
\bibitem{WilsonKogut}
K. G. Wilson, J. Kogut, ``The Renormalization Group and the $\epsilon$ Expansion'',
Phys. Rep. 12 (1974) 75.\newline https://doi.org/10.1016/0370-1573(74)90023-4

\bibitem{Brown}
  L. M. Brown (ed.), \emph{Renormalization From Lorentz to Landau (and Beyond)}, Springer, New York, 1993.

\bibitem{Hollowood}
  T. J. Hollowood, \emph{Renormalization Group and Fixed Points in Quantum Field Theory}, Springer, Heidelberg, 2013.

\bibitem{Haag}
  R. Haag, \emph{Local Quantum Physics}, Springer, Berlin, 1996.

\bibitem{Fraser} D. Fraser, \emph{Haag's Theorem and the Interpretation of Quantum Field Theories with Interactions}, Ph.D. thesis, U. of Pittsburgh, 2006. http://d-scholarship.pitt.edu/8260/.

\bibitem{EarmanFraser}
J. Earman, D. Fraser, ``Haag's Theorem and its Implications for the Foundations of Quantum Field Theory'', Erkenntnis 64 (2006) 305. https://doi.org/10.1007/s10670-005-5814-y

\bibitem{BuchholzEM}
  D. Buchholz, F. Ciolli, G. Ruzzi, E. Vasselli, ``The Universal C*-Algebra of the Electromagnetic Field'' \emph{Lett. Math. Phys.} \textbf{106} (2016) 269. https://doi.org/10.1007/s11005-015-0801-y

\bibitem{IZ}
C. Itzykson, J.--B. Zuber, \emph{Quantum Field Theory}, McGraw--Hill, New York, 1980.

\bibitem{Streater}
R. F. Streater, ``Outline of axiomatic relativistic quantum field theory'', Rep. Prog. Phys. 38 (1975) 771.\newline https://doi.org/10.1088/0034-4885/38/7/001

\bibitem{Morgan1507}
P. Morgan, ``Multi–particle quantum fields for bound states and interactions'', arXiv:1507.08299
\end{thebibliography}
\end{document}